\documentclass[preprint,showpacs,preprintnumbers,amsmath,amssymb,aps,pra]{revtex4}
\usepackage{amssymb,amsmath,graphics,epsfig,amssymb,rotating}
\usepackage{color}
\usepackage{wrapfig}

\begin{document}

\title{{\bf Electric field excitation suppression in cold atoms}}
\author{Jianing Han \footnote{jhan@southalabama.edu}, Juliet Mitchell, Morgan Umstead} 
\affiliation{Physics Department, University of South Alabama, Mobile, AL 36688, USA}

\date{\today}
\begin{abstract}
\label{abst}  
In this article, the atom excitation suppression is studied in two ways. The first way of exploring the excitation suppression is by an external DC electric field. The second way is to study the excitation suppression caused by an electric field generated by free charges, which are created by ionizing atoms. This suppression is called Coulomb blockade. Here the Coulomb forces are created by ions through ionizing atoms by a UV laser. The theory shows that the interaction, which causes the suppression, is primarily caused by charge-dipole interactions. Here the charge is the ion, and the dipole is an atom. In this experiment, we use $^{85}$Rb atoms. The valence electron and the ion core are the two poles of an electric dipole. The interaction potential energy between the ion and the atom is proportional to $\frac{1}{R^2}$, and the frequency shift caused by this interaction is proportional to $\frac{1}{R^4}$, where $R$ is the distance between the ion and the dipole considered. 
This research can be used for quantum information storage, remote control, creating hot plasmas using cold atoms, as well as electronic devices.  
\end{abstract}

\maketitle
\section{Introduction}
We first study the ground state excitation suppression caused by the Stark shift of the atoms by adding an external DC electric field. We then study the Coulomb field effect on the excitation suppression. This can be also called Coulomb blockade. The Coulomb blockade was originally studied in electronic devices. The basic idea is that due to the electrons inside the electronic device, the conductivity of the device is affected \cite{Grabert, Mec, Dmitry}. In this article, we study the Coulomb blockade in a  cold dilute atomic gas. We purposely create charges inside the atomic gas, and we then study the energy shift caused by those charges, which in turn affects the conductivity of such gasses.

Very similar to the dipole-blockade studied in Rydberg gases, here the energy shift is caused by charges instead of dipoles. Dipole-blockade, the excitation suppression due to the energy shift caused by neighboring dipoles, has been proposed and experimentally studied due to its potential applications as quantum gates \cite{Haroche, Jaksch, Lukin, Martin, Tong, Singer, Heidemann, Liebish, Walker1, Pierre}. The interactions in dipole-blockade are caused by dipole-dipole interactions. Here each dipole is a Rydberg atom since Rydberg atoms have large electric dipole moments. The excited electron and the ion core in a Rydberg atom \cite{Gallagher4, Dunning} are the two poles of an electric dipole. In this article, we study the Coulomb blockade, or charge-dipole interactions, in the lower level atoms. Here we treat a lower level atom, a $5^2P_{3/2}$ atom, as a dipole. Those atoms do not have permanent dipoles; however, under an external electric field, such atoms will be polarized, or an induced dipole will be created. The advantage of Coulomb blockade over the dipole blockade is that it can work in an extensive range due to the stronger interaction strength between charges and dipoles. Specifically, we create ions and study ion-atom interactions or charge-dipole interactions. Such states can not only be used for quantum gates but also quantum information sciences. For example, this can be used to store quantum information. Moreover, the research reported can be used to create plasmas with a particular temperature \cite{Killian, Gallagher,Li,Tanner}.

This article is arranged in the following way: two ways of suppressing ground state excitation are described separately. A conclusion is given at the end. 

\section{DC field suppression}

\begin{figure}[tpb]
\centering
\includegraphics[width=3.0 cm,angle= 0 ,height=3.0 cm]{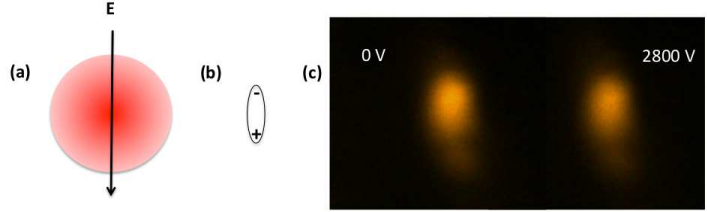}
\caption{ (a) The MOT at the center of the vacuum chamber and the electric field is added across the MOT as shown by {\bf E}. (b) The atoms in the MOT is polarized in the electric field. (c) Two images taken at a zero field and 1100 V/cm or 2800 V on two pairs of rods which are 1 inch apart. }
\label{schematic2}
\end{figure}

The experimental apparatus is very similar to experimental setups reported earlier \cite{HanMOT}. A Magneto-Optical Trap (MOT) is created by three pairs of retro-reflected beams. A repump laser is intersected with the three pairs of beams at the center of the vacuum chamber. An electric field is created between two pairs of rods. Four rods are located at the four corners of a square with the MOT at the center of the square \cite{Wenhui}. The side length of the square is one inch. Two of the rods are connected, and a high voltage is added on this pair of rods. The other pair is connected and grounded. Fig. \ref{schematic2}(a) is the schematic plot of the MOT, and the electric field created. Fig. \ref{schematic2}(b) shows one atom in the MOT polarized in the electric field direction. The data is taken by taking the images of the MOT. Fig. \ref{schematic2}(c) shows two images taken at fields 0 V/cm and 1100 V/cm respectively (or 0 V and 2800 V on one pair of rods and the other pair is grounded). It is shown that at a higher voltage or a higher field, the image intensity is lower, which indicates a lower number of excited atoms, or excitation suppression. The temperature of the atoms in the MOT is about 300 $\mu$K. We take the images of the MOT by a stingray camera at a 5 Hz repetition rate through a Vimba Viewer program. The images are then stored in a computer, integrated, and processed. The relative number of atoms is calculated by integrating the MOT image files. In this experiment, we vary the electric field strength and monitor the relative number of atoms in the MOT.

\begin{figure}[tpb]
\centering
\includegraphics[width=6 cm,angle= 0 ,height=6 cm]{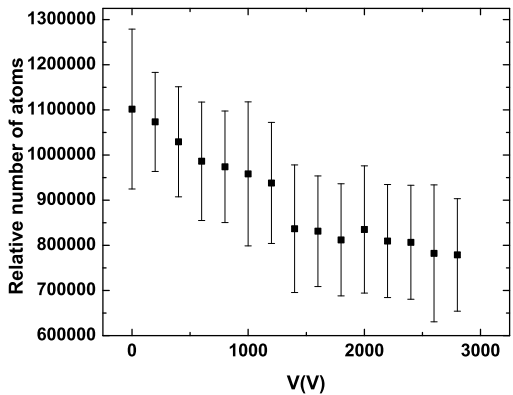}
\caption{ The relative number of atoms in the MOT vs. the voltage added on the electric field rods.}
\label{NvsVf}
\end{figure}
Fig. \ref{NvsVf} shows the relative number of atoms in the MOT vs. the voltage added. The relative number of atoms in the MOT is calculated by averaging at least two points taken at the same voltage. The error is the standard deviation of the points considered. The relative number of atoms is close to the actual number of atoms, and all the data points in Fig. \ref{NvsVf} are taken on the same day. The total number of atoms varies from day to day. Therefore, a relative number of atoms is used in this article. The field is roughly estimated by $E=\frac{V}{d}$, where V is the voltage added on two of the four rods located at the four corners of a square, and d is about one inch, the distance between two neighboring rods. The data taken on different days show the same trend, and by switching fields on and off, clear suppression has been repeatedly observed.

We have calculated the Stark effect for the trapping transition and repump transition. The Hamiltonian that we used to calculate the Stark effect is the following \cite{Adam}:
\begin{equation}
\begin{split}
H_{Stark}=-\frac{1}{2}\alpha_0E^2_z-\frac{1}{2}\alpha_2E^2_z\frac{3J^2_z-J(J+1)}{J(2J-1)}.
\end{split}
\label{Hstark}
\end{equation}
We define the electric field direction as the z direction, and the magnitude of the electric field added is $E_z$. $\alpha_0$ and $\alpha_2$ are the scalar and tensor polarizabilities. $J$ and $J_z$ are the total electron angular momentum state and the projection of the total electron angular momentum state in the z axis.  In this calculation, the ground state scalar polarizability that we used is $\alpha_0=0.0794h$ Hz/(V/cm)$^2$ for 5$^2$S$_{1/2}$ states. The 5$^2$P$_{3/2}$ scalar polarizability is $\alpha_0=0.2134h$ Hz/(V/cm)$^2$ and the 5$^2$P$_{3/2}$ tensor polarizability is $\alpha_2=-0.0406h$ Hz/(V/cm)$^2$, where $h$ is the Planck's constant \cite{Adam}.

\begin{figure}[tpb]
\centering
\includegraphics[width=3.2 cm,angle= 0 ,height=3.2 cm]{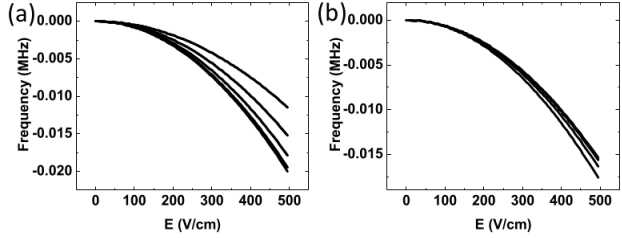}
\caption{ (a) The Frequency shift between the 5$^2P_{3/2}$ F=4 state and the 5$^2S_{1/2}$ F=3 state under an electric field, or the Stark effect of the trapping transition of the $^{85}$Rb atoms. (b) The Frequency shift between the 5$^2P_{3/2}$ F=3 state and the 5$^2S_{1/2}$ F=2 state under an electric field, or the Stark effect of the repump transition of the $^{85}$Rb atoms. In both plots, we shift the zero field energy difference to zero.}
\label{stark}
\end{figure}

Fig. \ref{stark} shows the energy shift caused by the Stark effect. In this plot, we calculated the energy difference between the 5$^2P_{3/2}$ F=4 state and the 5$^2S_{1/2}$ F=3 state, or the trapping transition as shown in Fig. \ref{stark}(a), relative to the energy difference between those two states at  zero electric fields, $E$. 
\begin{equation}
E=(E_{5^2P_{3/2},F=4,DC}-E_{5^2S_{1/2},F=3,DC})-(E_{5^2P_{3/2},F=4}-E_{5^2S_{1/2},F=3}).
\label{E_level}
\end{equation}
where $E_{5^2P_{3/2},F=4}$ is the energy of the $5^2P_{3/2},F=4$ state. $E_{5^2P_{3/2},F=4,DC}$ is the energy of the same state under an external DC electric field. The different energy levels correspond to the different $F_z$ states, where $F_z$ is the projection of the total atomic angular momentum, $F=I+J$, in z direction. $I$ is the nuclear angular momentum. For example, the top state is the energy difference between $5^2P_{3/2},F=4, F_z=4$ and $5^2S_{1/2},F=3$ state calculated from Eq. \eqref{E_level}. Since the different $F_z$ states of $5^2S_{1/2},F=3$ at the field strength considered are degenerate, therefore, no $F_z$ state is specified for $5^2S_{1/2},F=3$. Similar to the trapping transition, the corresponding repump transition is also calculated as shown in Fig. \ref{stark}(b). The different levels again correspond to different $F_z$ states. It is shown that the span (the span on the right side of Fig. \ref{stark}(a) and \ref{stark}(b)) of the trapping transition at about 500 V/cm, which corresponds to the voltage 1270 V on the two pairs of rods in our case, is about 10 kHz, and the span of the repump transition at the same voltage is about 3 kHz. The line width of the trapping laser or repump laser is about 1 MHz. The fact that the frequency span at 500 V/cm is much less than the laser linewidth and more than 10$\%$ suppression has been observed for this state as shown in Fig. \ref{NvsVf}, which indicates that the polarization of atoms under an electric field may contribute to the population suppression. 

\section{Coulomb Blockade}
\begin{figure}[tpb]
\centering
\includegraphics[width=6 cm,angle= 0 ,height=6 cm]{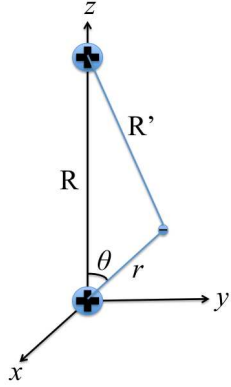}
\caption{ An ion at the top and an atom at the origin of the Cartesian coordinate are separated by a distance R.}
\label{ion_dipole}
\end{figure}

To study Coulomb blockade, we ionize some atoms in the MOT, and we examine the population suppression. Here is the brief theory. We calculate the frequency shift of the energy level of one atom in the presence of an ion. The energy shifts of an atom are caused by ion-dipole interactions. Here we treat the atom under the Coulomb field as an induced dipole. A valence electron and an ion core are the two poles of an electric dipole. As shown in Fig. \ref{ion_dipole}, an atom and a free ion are separated by a distance R. The ion core of the atom is at the origin of the x, y, and z axis. The radius of this atom is r. The distance between the electron of the atom and the free ion is R'. The angle $\theta$ is the angle between R and r. In this article, we assume the radius of the atoms, r, is much smaller than the spacing between the two ions, R. We apply the Born-Oppenheimer approximation, that is, the wave functions of the ion cores and electrons can be separated. We further assume that the ions are at rest. Then the Hamiltonian of the three-body system shown in Fig. \ref{ion_dipole} can be written as:
\begin{equation}
\begin{split}
H=-\frac{\hslash ^2}{2m}\triangledown ^2+V,
\end{split}
\label{H}
\end{equation}
where $m$ is the effective electron mass; $\hslash =\frac{h}{2\pi}$ and $h$ is the Planck's constant. The potential energy V in Eq. \eqref{H} can be written as:
\begin{equation}
\begin{split}
 V=\frac{e^2}{4\pi \epsilon_0}(-\frac{1}{r}-\frac{1}{R'}+\frac{1}{R}),
\end{split}
\label{V}
\end{equation}
where $e$ is the electron charge; and $\epsilon_0$ is the permittivity of free space. 
Fig. \ref{ion_dipole} shows that 
\begin{equation}
\begin{split}
 \frac{1}{R'}=\frac{1}{\sqrt{(R-rCos(\theta))^2+r^2Sin^2(\theta)}}.
\end{split}
\label{1overR}
\end{equation}
The above equation can be also derived from the law of Cosine. If $R>>r$, we can do the Taylor expansion for $\frac{1}{R'}$ at R. The first few orders are shown below:
\begin{equation}
\begin{split}
 \frac{1}{R'}\approx \frac{1}{R}+\frac{rCos(\theta)}{R^2}+\frac{r^2(3Cos^2(\theta)-1)}{2R^3}+\cdot\cdot\cdot\cdot\cdot\cdot
\end{split}
\label{1overR2}
\end{equation}
If we keep the first two terms, the Hamiltonian can be rewriten as
\begin{equation}
\begin{split}
H\approx -\frac{\hslash ^2}{2m}\triangledown ^2+\frac{e^2}{4\pi \epsilon_0}(-\frac{1}{r}-\frac{1}{R}-\frac{rCos(\theta)}{R^2}+\frac{1}{R}),
\end{split}
\label{H2}
\end{equation}
or
\begin{equation}
\begin{split}
H\approx -\frac{\hslash ^2}{2m}\triangledown ^2-\frac{e^2}{4\pi \epsilon_0}\frac{1}{r}-\frac{e^2}{4\pi \epsilon_0}\frac{rCos(\theta)}{R^2}.
\end{split}
\label{H3}
\end{equation}
Notice that the combination of the first two terms is the Hamiltonian of a Hydrogen atom, and the last term is the perturbation. This Hamiltonian can be solved by diagonalizing the matrix composed of the eigen functions of a Hydrogen atom. In this article, we simplify this problem by taking advantage of the known Stark effect calculations as shown in Eq. \eqref{Hstark} \cite{Adam}. Since $rCos(\theta)=z$, we just need to replace the external electric field by the Coulomb field: $\frac{e}{4\pi \epsilon_0R^2}$. More specifically, the interaction energy between an ion and an atom can be written as:

\begin{equation}
\begin{split}
H_{ion-dipole}=-\frac{1}{2}\alpha_0(\frac{e}{4\pi \epsilon_0R^2})^2-\frac{1}{2}\alpha_2(\frac{e}{4\pi \epsilon_0R^2})^2\frac{3J^2_z-J(J+1)}{J(2J-1)}.
\end{split}
\label{Hion}
\end{equation}
We use the same scalar and tensor polarizabilities, $\alpha_0$ and $\alpha_2$, described in the DC field suppression session.

\begin{figure}[tpb]
\centering
\includegraphics[width=6 cm,angle= 0 ,height=6 cm]{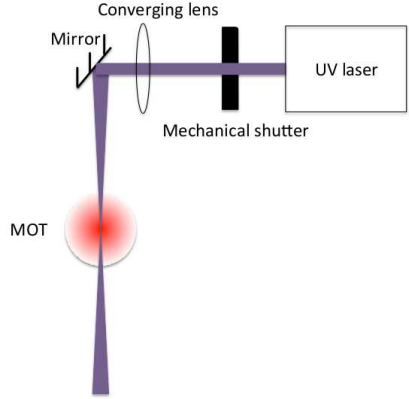}
\caption{ The schematic plot of this experiment.}
\label{schematic}
\end{figure}
Fig. \ref{schematic} shows the schematic plot of this experiment. Here we use a UV laser to ionize atoms because a UV laser can create electrons and ions with greater kinetic energies. It has been shown in previous literature \cite{Wenhui2} that the electrons will leave the MOT first and leave the ions behind, which is caused by the momentum conservation. Specifically, the electrons will move a lot faster than the ions once they are ionized, since the electron mass is much smaller than the mass of the ion core. Therefore, using a high energy laser, the pure ion environment will be achieved faster due to the greater kinetic energy of the electrons. A continuous UV diode laser produces a UV laser light at 405 nm. The power of this UV laser is about 100 mW, and the temperature of this UV laser is not stabilized. We would expect that the wavelength will drift slightly over time. However, the temperature is relatively stable at the time that the data was taken, since we typically wait for some time to let the system stabilize. The laser beam will then pass through a mechanical shutter. The shutter's turning on speed is 10 ms. After passing a converging lens, the laser beam is focused on the MOT (Magneto-Optical Trap) held in a vacuum chamber.

\begin{figure}[tpb]
\centering
\includegraphics[width=12 cm,angle= 0 ,height=12 cm]{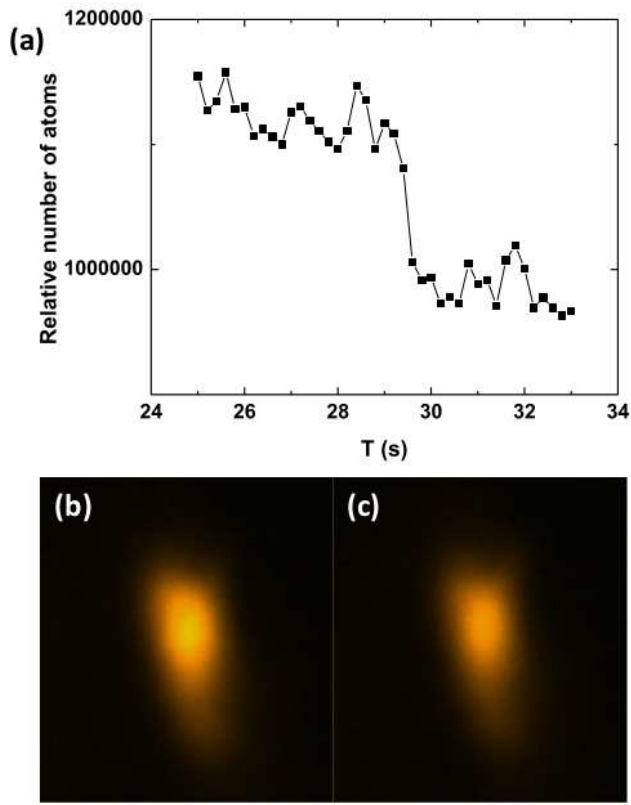}
\caption{ (a) The relative number of atoms as a function of time. The UV light was turned on at about 29 s as shown in this plot. (b) The image was taken at about 28.4 s as shown in (a). (c) The image was taken at about 32.4 s as shown in (a).}
\label{plasma_suppression}
\end{figure}
Fig. \ref{plasma_suppression} shows the plasma suppression data and images. Fig. \ref{plasma_suppression}(b) and \ref{plasma_suppression}(c) show two images taken without the ions (Fig. \ref{plasma_suppression}(b)) and with ions (Fig. \ref{plasma_suppression}(c)). It is shown that fewer atoms are excited with the ions, or excitation suppression is observed. The vertical axis is the relative number of cold atoms, and the horizontal axis is the time. It is shown that the percent suppression is about 15$\%$. In this experiment, the 405 nm UV light is turned on at about 29 s by opening the shutter in front of the UV laser (this set of data is one portion of a larger set of data, and the trend reported here is repeatable). The images are continuously taken at a 5 Hz repetition rate. Since the data were continuously taken, no error bars are given in this plot. The time is estimated from the image taken rate. This rate is confirmed by the shutter switching rate. For example, if the shutter switching rate is set at 0.1 Hz, then within one shutter switching cycle, 50 images will be taken. It is shown that the MOT density gradually decreases when the UV laser was shined on the MOT. The MOT unloading time by blocking one of the trapping beams is much less than 0.2 s. The spacing between neighboring atoms is estimated to be 2 $\mu$m, and the distribution of the atoms is random, or non-uniform.  

\begin{figure}[tpb]
\centering
\includegraphics[width=3.5 cm,angle= 0 ,height=3.5 cm]{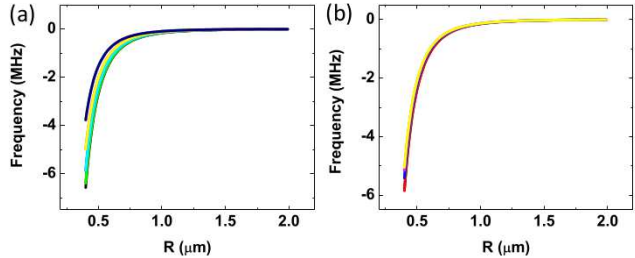}
\caption{ (a) The trapping transition shift due to ions. (b) The repump transition shift due to ions.}
\label{Energy_level}
\end{figure}

Fig. \ref{Energy_level} shows the energy shift of the trapping transition and repump transition under an electric field produced by an ion as shown in Fig. \ref{ion_dipole}. For example, Fig. \ref{Energy_level}(a) is calculated in the following way. We first calculate the energy difference between the $5^2P_{3/2}$ F=4 level and the $5^2S_{1/2}$ F=3 level under an ion field. We then subtract this energy difference under an ion field by the energy difference between those two states under zero electric field. Another way of expressing this is 
\begin{equation}
E=(E_{5^2P_{3/2},F=4,ion}-E_{5^2S_{1/2},F=3,ion})-(E_{5^2P_{3/2},F=4}-E_{5^2S_{1/2},F=3}).
\label{E_level}
\end{equation}
For example, the difference between $E_{5^2P_{3/2},F=4,ion}$ and $E_{5^2P_{3/2},F=4}$ is that $E_{5^2P_{3/2},F=4,ion}$ is the energy under an ion field, while $E_{5^2P_{3/2},F=4}$ is the energy of the same level under zero field. Similar to the DC field case, the different levels correspond to different $F_z$ states. It is shown that at about 0.8 $\mu$m, the maximum frequency shift caused by the ion field is about 0.5 MHz, which is comparable to the laser linewidth that we used for this experiment. Considering the non-uniform distribution of the atoms in the MOT, the theory is reasonable to explain the 15$\%$ suppression of the experimental data. Here we assume that the $\frac{1}{R^4}$ dependence dominates, and other $R$ dependence is still under investigation. The question remains is that: would the excitation from $5^2P_{3/2}$ F=4 to the ionized state contributes the decrease of the population as shown in Fig. \ref{plasma_suppression}(a)? To answer this question, we have calculated the oscillator strength between the $5^2P_{3/2}$ F=4 and $5^2S_{1/2}$ F=3 as well as the oscillator strength between the $5^2P_{3/2}$ F=4 and a highly excited state, such as n=1000. It turns out that the latter, the oscillator strength between the $5^2P_{3/2}$ F=4 and a highly excited state (n=1000), is much less than one percent of the former, the oscillator strength between the $5^2P_{3/2}$ F=4 and $5^2S_{1/2}$ F=3. We would expect the oscillator strength between $5^2P_{3/2}$ F=4 and a free state or an ionized state is even lower. Therefore, this confirms that the population suppression is caused by the electric field by the free charge produced by ionization.

\section{Conclusion}

We have shown two types of excitation suppression. We first apply an external DC electric field to suppress the ground state excitation. We then studied the suppression caused by a Coulomb field. We have shown how to calculate the frequency shift caused by the Coulomb field from an ion by taking advantage of the known Stark effect calculations. We have found that the frequency shift is about half a MHz at R=0.8 $\mu$m, where R is the distance between the ion and the atom. Moreover, we have experimentally observed 15$\%$ excitation suppression caused by the ions produced by a UV laser at the average internuclear spacing 2 $\mu$m, which can be supported by the theory reported.

\section{Acknowledgement}

It is a pleasure to acknowledge very helpful discussions with Drs. T. F. Gallagher, G. Bo, and 
J. Sanders.

\end{document}